\documentclass[a4paper]{article}
\usepackage[utf8]{inputenc}
\usepackage{amsmath,amsfonts,amssymb,amstext,latexsym}
\usepackage{authblk}

\usepackage{graphicx}
\usepackage{bm}
\usepackage{url}

\usepackage[a4paper,left=2cm,right=2cm,top=1.5cm,bottom=1.5cm]{geometry}

\renewcommand{\vec}{\bm}

\newcommand{\be}{\begin{equation}}
\newcommand{\ee}{\end{equation}}
\newcommand{\bea}{\begin{eqnarray} \nonumber }%
\newcommand{\eea}{\end{eqnarray}}
\newcommand{\bi}{\begin{itemize}}
\newcommand{\ei}{\end{itemize}}

\newcommand{\cRM}[1]{\MakeUppercase{\romannumeral #1}} 
\newcommand{\cRm}[1]{\textsc{\romannumeral #1}} 

\newcommand{\dd}{\mathrm{d}}
\newcommand{\e}{\mathrm{e}}
\newcommand{\ii}{\mathrm{i}}
\newcommand{\m}{\mathrm{m}}

\newcommand{\hbe}{\hbar_\mathrm{exp}}

\newcommand{\Wconc}{W_\mathrm{conc}}
\newcommand{\vpconc}{\vec{p}_\mathrm{conc}}
\newcommand{\Econc}{E_\mathrm{conc}}

\title{Effective gravity and effective quantum equations in a system inspired by walking droplets experiments}
\author{Christian Borghesi \thanks{christian.borghesi@protonmail.com}\\
\normalsize{Lyc\'ee Fran\c{c}ais de Vienne, Liechtensteinstra{\ss}e 37a, 1090 Wien, Austria}
}
\date{}

\begin{document}
\maketitle

\begin{abstract}
In this paper we suggest a macroscopic toy system in which a potential-like energy is generated by a non-uniform pulsation of the medium ({\it i.e.} pulsation of transverse standing oscillations that the elastic medium of the system tends to support at each point). This system is inspired by walking droplets experiments with submerged barriers. We first show that a Poincaré-Lorentz covariant formalization of the system causes inconsistency and contradiction. The contradiction is solved by using a general covariant formulation and by assuming a relation between the metric associated with the elastic medium and the pulsation of the medium. (Calculations are performed in a Newtonian-like metric, constant in time). We find ($i$) an effective Schrödinger equation with external potential, ($ii$) an effective de Broglie-Bohm guidance formula and ($iii$) an energy of the `particle' which has a direct counterpart in general relativity as well as in quantum mechanics. We analyze the wave and the `particle' in an effective free fall and with a harmonic potential. This potential-like energy is an effective gravitational potential, rooted in the pulsation of the medium at each point. The latter, also conceivable as a natural clock, makes easy to understand why proper time varies from place to place.
\end{abstract}


\section*{Introduction}
Walking droplets experiments, in which droplets bounce and `walk' on a vibrating liquid substrate (initiated in~\cite{yc_walking05}, see \cite{bush_review15} for a review) have shown for the first time that classical and macroscopic systems exhibit quantum-like phenomena. The fact that droplets are guided by the wave that they have generated is reminiscent of the {\it pilot wave} suggested by de Broglie~\cite{ldb_ondeguidee1927,solvay_conf1927} (see \cite{yc_wavepartduality11,bush_review15,bush_phystoday16} for a discussion in this context) and, more generally, of the {\it double solution} program~\cite{ldb_tentative,ldb_interpretation1987} (see~\cite{fargue,double_sol_90ans} for current and well-thought overviews). Nevertheless, it seems tricky to mathematically formalize walking droplet problems in order to obtain equations close to those found in corresponding quantum systems. To deal with more convenient systems, from a mathematical point of view, we have recently suggested a classical and macroscopic system~\cite{masselotte} -- also inspired by an experiment with a sliding bead on a vibrating string~\cite{yc_sao1999} -- and, with more success, the model system~\cite{conc} that we describe below.

This macroscopic toy system consists of ($i$) an elastic medium, which carries transverse waves governed by a Klein-Gordon-like equation, and ($ii$) one high elastic medium density, considered as a point of mass $m_0$, called concretion~\footnote{The latter idea came from a comment of Poincaré according to whom matter could be considered as aether's hole -- {\it ``[...] si bien que l'on pourrait dire~: il n'y a pas de matière, il n'y a que des trous dans l'éther''}~\cite{poincare_meca-nouvelle} and commented in~\cite{pierseaux_RR}.}. This system is invariant by Lorentz-Poincaré transformations specific to the elastic medium (where the velocity of surface waves plays the role of the speed of light in special relativity). This approach was very encouraging. For instance it has been found: ($i$) a strictly analogous free Schrödinger equation; ($ii$) a covariant guidance formula (which leads to the effective de Broglie-Bohm guidance formula in the low-velocity approximation); and ($iii$) the energy and momentum of the `particle' concretion have a direct counterpart in special relativity as well as in quantum mechanics. However these results only concern the special case without external potential. It is then natural to wonder how an external potential in this system could be generated and, afterward, whether previous results hold in presence of an external potential. The aim of this study is to answer these questions.

The paper is organized as follows. In Section~\ref{s1} we present the macroscopic toy system studied in this paper, which is a direct continuation of the one in~\cite{conc} apart from non-uniform {\it pulsation of the medium} (PM). An effective Schrödinger equation with external potential is obtained -- provided that some conditions and approximations are fulfilled -- with an external potential. Nevertheless this formulation results in a contradiction in the system, in particular for the behavior of the `particle' concretion. To overcome this contradiction, we develop a general covariant formulation of a system in Section~\ref{sacc}. (Almost all calculations are actually performed under the limit of an effective Newtonian metric.) This formulation allows us to solve the contradiction encountered in Section~\ref{s1} for a system with non-uniform PM. Finally, we study the free-fall example and an effective harmonic potential.

\section{Modifying pulsations of the medium \label{s1}}

\subsection{Inhomogeneous medium in walking droplets experiments}
By using a container with different fluid depths in walking droplets experiments, authors have studied tunnel-like effects~\cite{yc_tunnel09,influence_depth} (see also \cite{bush_tunneling,self-propulsion} for theoretical investigations) and non-specular reflection of walking droplets~\cite{gpucci_reflection}. A barrier with a different thickness in the vibrating cell changes properties of the system at the location of the barrier, in particular the threshold for Faraday instability. This allows them to generate a kind of potential barrier acting both on walkers and on surface waves. This way to experimentally generate a potential barrier has inspired us in this paper. We suggest that an inhomogeneous property of the elastic medium in our macroscopic toy system could generate an effective potential; and this property is the PM at each point, {\it i.e.} pulsation of transverse standing oscillations that the elastic medium tends to support at each point.

Note that Fig.~1 in~\cite{yc_tunnel09} shows that the submerged barrier in the cell increases the Faraday threshold acceleration. This means, when the forcing vibration amplitude is uniform and fixed, that the Faraday pulsation (if it exists) at the location of the barrier (with fluid depth $h_1$) should be higher than in the container outside the barrier (with fluid depth $h_0$, such that $h_0>h_1$). In other words, for a container with a uniform fluid depth $h_1$, and for the same forcing vibration amplitude, the forcing pulsation (then the
Faraday pulsation, $\Omega_\mathrm{F 1}$) of the cell in order to obtain sustained Faraday waves should be higher than the one with $h_0$ ($\Omega_\mathrm{F 0}$). This qualitatively corresponds in our system to an increasing potential-like energy at the location of the submerged barrier (where $\Omega_\mathrm{F 1}>\Omega_\mathrm{F 0}$).

Let us now briefly present our macroscopic toy system with uniform PM and explain how it is inspired by walking droplets experiments (see~\cite{conc} for a more detailed presentation).

\subsection{Brief recall of the system with homogeneous PM \label{ss1_syst}}
The main goal of the devised systems in~\cite{masselotte,conc} is to deal ($i$) with main characteristics of walking droplets experiments and ($ii$) with a convenient formalism to identify quantum analogies (and differences). The vibrating liquid carrying transverse waves ({\it i.e.} the interface height $h(\vec{r},t)$) becomes in our system an elastic medium carrying the transverse wave $\varphi(\vec{r},t)$ (which is also real-valued). This allows us to deal with traveling transverse waves {\it à la} d'Alembert instead of (dispersive) traveling capillary waves. In these experiments, quantum-like behaviors are associated with a specific property of the system: (since the system is near the Faraday instability threshold) each transverse perturbation at the surface of the bath tends to generate a harmonic oscillation at the Faraday pulsation ($\Omega_\mathrm{F}$) at the location of the perturbation. (This is related to the {\it memory} of the system~\cite{yc_JFM11}.) This property is mimicked in our system in the following way: due to a quadratic/harmonic potential, any element of the elastic medium tends to support transverse harmonic oscillations at pulsation $\Omega_\m$ ({\it i.e.} the frequency of these oscillations should be $\frac{\Omega_\m}{2\pi}$). Regarding the (stable) walking droplet free to walk upon the liquid bath, this becomes in our system a (stable) `particle', called concretion, free to slide on the elastic medium~\footnote{In~\cite{yc_sao1999} the authors perform and analyze a sliding bead on a vibrating string experiment. This has inspired us and has also been investigated by means of a Lagrangian formulation~\cite{masselotte}.}. With some similarities with a droplet, the concretion is considered as a high mass-density of the elastic medium itself and has the same properties as the homogeneous elastic medium per unit mass. (In~\cite{masselotte}, the `particle' was an exogenous bead oscillator, but analogies with quantum mechanics were less satisfying than those obtained with the endogenous concretion~\cite{conc}.) From this point of view, our classical system is wave monistic.

In this toy system, all results are deduced from a Lagrangian formalism which is invariant by Lorentz-Poincaré transformations ({\it e.g.}~\cite{feynman_em2} \S 25) specific to the elastic medium (where the velocity of surface waves plays the same role as the speed of light in special relativity). A wave equation is then deduced: a Klein-Gordon-like equation, in which the concretion can be a source of waves. The system is governed by equations similar to those in quantum mechanics when the concretion is not a source of waves, {\it i.e.} there is no longer back-reaction of the concretion on the wave. This special case was called {\it symbiosis} between the wave and the `particle' concretion. This state, related to an intimate harmony between the wave and the concretion, presents some analogies with walking droplets experiments: stable orbits of walkers (which present some analogies with quantum states), resulting from self-organized phenomena between walkers and the wave at the surface of the bath, are such that walkers do not excite (or very slightly) the main eigenmode of the global wave~\cite{yc_SO-eingenstates14}. This fact is even more significant as a walker generates a wave at each bounce.

As shown in previous works, the Lagrangian density of the system reads
\be \label{econc_l}
\mathcal{L} = \frac{1}{2}\, \mathcal{T}\,\left(1\,+ \, \frac{\rho_0(\vec{r},t)}{\mu_0} \right) \left[\partial_\mu \varphi \, \partial^\mu \varphi  \,-\, \frac{\Omega_\m^2}{c_\m^2} \varphi^2 \right] \,,
\ee 
where $\mathcal{T}$ denotes a `tension' of the elastic medium, $\mu_0$ its mass per element of volume, $c_\m$ the propagation speed of the wave (such that $\mathcal{T} = \mu_0\,c_\m^2$), $\Omega_\m$ the {\it reference pulsation of the medium} and $\rho_0$ the very high elastic medium density corresponding to the concretion of mass $m_0$ at position $\vec{\xi}$ at time $t$ in a reference frame $\mathcal{R}$ ({\it i.e.} $\rho_0(\vec{r}_0,t_0) = m_0\,\delta(\vec{r}_0-\vec{\xi}_0)$ in its proper reference frame $\mathcal{R}_0$, {\it e.g.}~\cite{landau_th_ch} \S 28, in which $\delta$ denotes the Dirac delta function). Here, $\partial_\mu \varphi \, \partial^\mu \varphi$ means $(\frac{1}{c_\m}\frac{\partial\varphi}{\partial t})^2 - (\vec{\nabla}\varphi)^2$. 

Let us now briefly comment this macroscopic toy system. The concretion is assumed to be a stable particle (like walkers in usual walkers' dynamics) and can be seen as a simplification of more complicated phenomena (for instance peaked solitons, due to a non-linear self-focusing potential of gravitational nature, developed in very interesting studies~\cite{durt_1,durt_2} inspired by walking droplets experiments and de Broglie's double solution program). In contrast to de Broglie's double solution program (in which particles consist of a peaked concentration of energy), the `particle' concretion is depicted by a very high elastic medium density rather than a very high amplitude of the wave. Moreover, the transverse wave $\varphi$ -- continuous at the location of the concretion -- also plays the role of a ``pilot wave". Finally, this system can be interpreted as a simplification (with the modification of the particle's description) of de Broglie's double solution program in a classical system. It is clear by the way from the Lagrangian formalism presented in this section that the concretion is treated as a particle; in this perspective, the model is more reminiscent of Bohm's dynamics than of de Broglie's double solution. However it could be interpreted as a limiting case of a wave monistic model, still to write.

According to the nature of the concretion ({\it i.e.} a point-like high density elastic medium), we seek in the following an external potential which acts in the same way on the elastic medium and on the concretion (but naturally $\rho_0/\mu_0$ times more). This facilitates to retrieve a potential-like energy able to act on the concretion, from a wave equation valid at any point.

\subsection{Framework}
Up to now PM was uniform and constant in time -- and equal to $\Omega_\m$. We assume in this section that PM is no longer uniform. But for the sake of simplicity we assume that these pulsations are constant in time. The PM at position $\vec{r}$ then reads:
\be 
\Omega_\m'(\vec{r}) = \Omega_\m \,+\, \omega_\m(\vec{r}) \,.
\ee
This means that the elastic medium at position $\vec{r}$ tends to support a transverse standing oscillation at pulsation $\Omega_\m'(\vec{r})$. 

We assume through this paper that $\omega_\m(\vec{r})$ ($i$) can be positive or negative, ($ii$) has a magnitude much lower than $\Omega_\m$ ({\it i.e.} $|\omega_\m|\ll \Omega_\m$) and ($iii$) is very smooth comparatively to characteristics related to $\Omega_\m$ ({\it i.e.} $\omega_\m(\vec{r})$ has a characteristic length evolution much greater than $c_\m/\Omega_\m$).

\begin{figure}
\begin{minipage}[c]{0.6\linewidth}
\centering
\includegraphics[width=1 \columnwidth, clip=true]{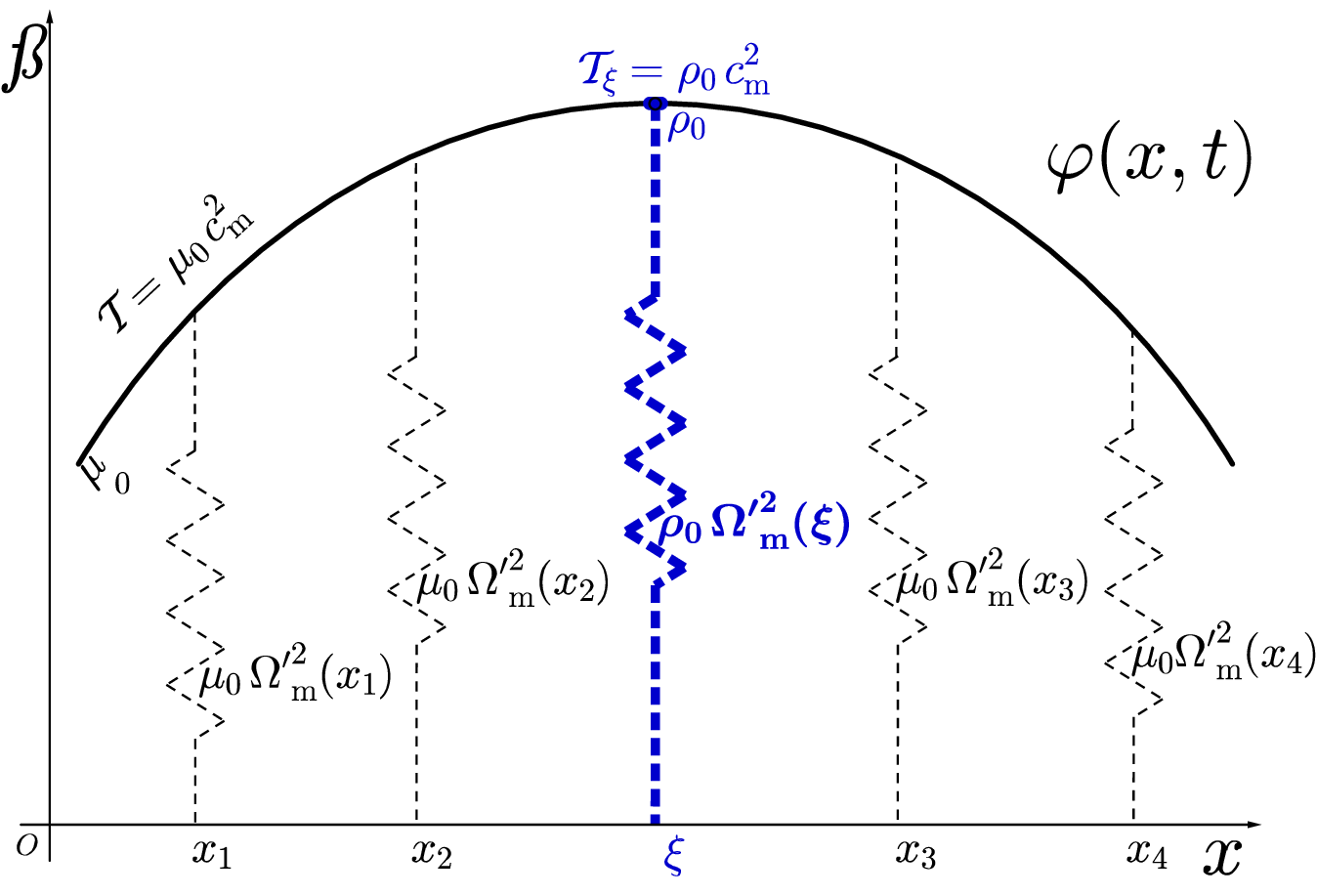}\hfill
\end{minipage}
\begin{minipage}[c]{0.35\linewidth}
\caption{Schematization of the theoretical system, here in a 1D elastic medium and in the proper reference frame of the concretion. (Transverse oscillations are directed along the {\it eszett} axis, $(O\ss)$.) PM, $\Omega_\m'(x)$, is written for various positions. Transverse harmonic potentials of the medium are depicted by springs, whose stiffness per element of length is indicated and use $\Omega_\m'^2(x)$. The point-like high elastic medium density, {\it i.e.} the concretion (in blue), has the same properties per unit mass as the homogeneous (in mass) elastic medium.}
\end{minipage}
\label{fig1}
\end{figure}

Non-uniform PM, $\Omega_\m'(\vec{r})$ instead of $\Omega_\m$, constitutes the only difference from the macroscopic toy system suggested in~\cite{conc}. (See Fig.~\ref{fig1} for a schematic representation of the system studied in this paper.) The Lagrangian density of the system then becomes
\be \label{e1_l}
\mathcal{L} = \frac{1}{2}\, \mathcal{T}\,\left(1\,+ \, \frac{\rho_0(\vec{r},t)}{\mu_0} \right) \left[ \partial_\mu \varphi \, \partial^\mu \varphi \, - \, \frac{\Omega_\m'^{2}(\vec{r})}{c_\m^2} \varphi^2 \right] \,. 
\ee
It is important to note that the pulsation of the medium, $\Omega_\m'(\vec{r})$, acts on the homogeneous elastic medium as well as on the concretion (but $\rho_0/\mu_0$ times more).

\subsection{Wave equation and symbiosis equation}
The wave equation results from the Euler-Lagrange equation for a scalar field (which stems from the principle of least action, when the wave field $\varphi$ is subjected to a small change $\delta\varphi$ while the 4-position of the `particle' concretion is not subjected to a small change), {\it i.e.} $\partial_\mu \frac{\partial \mathcal{L}}{\partial(\partial_\mu \varphi)} = \frac{\partial \mathcal{L}}{\partial \varphi}$. According to the Lagrangian density~(\ref{econc_l}) of the system, this leads to $\partial_\mu\left[(1+\rho_0/\mu_0)\partial^\mu \varphi\right] = -(1+\rho_0/\mu_0)\frac{\Omega_\m'^2}{c_\m^2}\,\varphi$; {\it i.e.}
\be \label{e1_ch}
\square_\m\varphi \,+\, \frac{\Omega_\m'^2(\vec{r})}{c_\m^2}\,\varphi 
 = -\,\frac{\rho_0}{\mu_0}\left(\square_\m\varphi \,+\, \frac{\Omega_\m'^2(\vec{r})}{c_\m^2}\,\varphi \right) \, - \, \frac{1}{\mu_0}\left[\partial_\mu(\rho_0)\,\partial^\mu \varphi \right] ,
\ee
where $\square_\m$ denotes the d'Alembert operator specific to the elastic medium ({\it i.e.} $\square_\m=\frac{1}{c_\m^2}\frac{\partial^2}{\partial t^2}-\Delta$, where $\Delta$ denotes the Laplace operator).

The concretion is no longer the source of the wave (a state called symbiosis between the wave and the concretion) when the two following conditions are fulfilled: ($i$) the wave $\varphi$ obeys a Klein-Gordon-like equation with non-uniform pulsations $\Omega_\m'(\vec{r})$ ({\it i.e.} $\square_\m\varphi \,+\, \frac{\Omega_\m'^2(\vec{r})}{c_\m^2}\,\varphi=0$) and ($ii$) the following equation,
\be \label{e1_symb}
\partial_\mu(\rho_0)\,\partial^\mu \varphi=0 \,,
\ee
called the symbiosis equation, is satisfied. (This equation reads $\frac{1}{c_\m^2}\,\frac{\partial \rho_0}{\partial t}\,\frac{\partial \varphi}{\partial t} - \vec{\nabla}\rho_0 \cdot \vec{\nabla}\varphi = 0$.) We will discuss this equation below.

\subsection{Effective Schrödinger equation with an external potential}
The derivation of the Schrödinger equation from the Klein-Gordon equation in the low-velocity approximation is mentioned by de Broglie (see {\it e.g.} \cite{ldb_tentative} \S \cRM{2}.7) and is well-known in the literature (see {\it e.g.} \cite{zee} \S \cRM{3}.5). In our classical context it is convenient to introduce the modulating wave, $\psi$, which modulates the `natural' (standing) wave of the medium without any `particle'. Using complex notation and the reference PM, $\Omega_\m$, the (complex-valued) modulating wave $\psi$ is defined from the (real-valued) transverse wave, $\varphi$, as
\be \label{epsi_def}
\varphi(\vec{r},t) = \mathrm{Re}\left[A~ \psi(\vec{r},t) ~ \e^{-\ii\,\Omega_\m\,t}  \right] \,,
\ee
where $\mathrm{Re}[\cdots]$ denotes the real part and $A$ an amplitude of the transverse oscillation. (Then, the modulating wave $\psi$ is dimensionless.) 

When the wave is in symbiosis with the concretion, the wave equation~(\ref{e1_ch}) in low-velocity and small $\omega_\m$ approximation leads to
\be \label{e1_schro0}
\ii\,\frac{\partial \psi}{\partial t} = -\,\frac{c_\m^2}{2\,\Omega_\m}\Delta\psi \:+\: \omega_\m\,\psi \,.
\ee
(Calculations are straightforward and very similar in our context to the ones detailed in the appendix of~\cite{conc} apart from inhomogeneous pulsations.) Note that what we call throughout this paper low-velocity and small-potential approximation is a first-order approximation with respect to $v^2/c_\m^2$ and $\omega_\m/\Omega_\m$.

Now, we use the coefficient 
\be \label{ehbe}
\hbe=\frac{m_0\,c_\m^2}{\Omega_\m} \,,
\ee
which is specific to the studied system. ($\hbe$ has been introduced in~\cite{masselotte} as a proportionality coefficient between wave characteristics and particle characteristics, but appears naturally in the energy of the concretion~\cite{conc}.) Thus, Eq.~(\ref{e1_schro0}) becomes:
\be \label{e1_schro}
\ii\,\hbe\,\frac{\partial \psi}{\partial t} = -\,\frac{\hbe^2}{2\,m_0}\,\Delta\psi \,+\, V\,\psi \,.
\ee
This equation has the same form as the Schrödinger equation with an external potential, $V$, acting on the concretion and defined as:
\be \label{e1_V}
V(\vec{r}) = \hbe\;\omega_\m(\vec{r}) \,.
\ee
We see below, in Section~\ref{ssge_major}, that this expression results naturally from a general covariant formalization of the system. At this point, expression~(\ref{e1_V}) shows that the potential-like energy acting on the concretion depends naturally on the two following points: ($i$) the mass $m_0$ of the concretion and ($ii$) a property of the elastic medium at position $\vec{r}$, namely the additional pulsation $\omega_\m(\vec{r})$.

\subsection{From the symbiosis equation to a contradiction of the system \label{ss1_contradiction}}
The symbiosis equation~(\ref{e1_symb}) as well as the mass continuity equation (related to the concretion) have the same expressions as in a system with uniform PM, $\Omega_\m'(\vec{r})=\Omega_\m$. Thus, the symbiosis equation associated with the conservation of the mass of the concretion leads to results similar to those of~\cite{conc} (cf. Eq.~(6) therein). (This equation leads to $\frac{\dd }{\dd t}\frac{1}{\sqrt{1-v^2/c_\m^2}} = 0$ as well as to a covariant guidance formula, $\frac{\vec{v}}{c_\m^2}\,\frac{\partial \varphi}{\partial t}(\vec{\xi},t) +  \vec{\nabla}\varphi(\vec{\xi},t) = \vec{0}$, where $\vec{v}=\frac{\dd \vec{\xi}}{\dd t}$ is the velocity of the concretion.) This implies that the velocity of the concretion remains constant in time. This is incoherent and wrong for the system suggested in this paper. Indeed, the velocity of the concretion under the action of an external potential-like $V$ should normally be able to change in time, instead of remaining constant in time. (We note also that the energy of the concretion, evaluated in the same way as in~\cite{conc} would again have been wrong.)

The toy system suggested in this section then exhibits an inconsistency. Something else in this system or in this formalization must be missing or wrong.

\section{General covariant formulation for a system with inhomogeneous pulsation of the medium \label{sacc}}
The previous contradiction is resolved by using a general covariant formulation, instead of the Lorentz-Poincaré covariant formulation (with respect to $c_\m$) of the previous Section. This allows us to deal with an heterogeneous system, while maintaining a covariant formulation. In general relativity ({\it e.g.}~\cite{landau_th_ch}) a gravitational field and the motion of a particle in a gravitational field can be described by the metric tensor (and the Newtonian metric is simplest metric used to describe the motion of a particle in the low-velocity and small gravitational field approximation). Based on these facts, we assume that the effective potential generated by non-uniform PM (as seen in the previous Section) and its action on the motion of a concretion (as it should be) can be described by a metric tensor associated with the elastic medium. Furthermore, this metric tensor of the elastic medium should be related to non-uniform PM (and this metric could be a Newtonian-like one in the low-velocity and small potential approximation). In short, we apply recipes from general relativity in order to deal with the concretion in an elastic medium with inhomogeneous PM.

Hence, firstly we study our system in curvilinear coordinates. Secondly we make an assumption which links PM to the metric tensor of the elastic medium. This allows us to write a guidance formula, a momentum and an energy of the concretion without contradiction. Finally we study the concretion in an effective free fall and in an effective harmonic potential.

\subsection{Framework}
Let a reference frame $\mathcal{R}$ in which spatio-temporal coordinates are written as $(\vec{r},t)$ or $x^\mu$. Throughout this paper $x^0=c_\m\,t$ and $x^i$ ($i=1,2,3$) denotes the spatial location $\vec{r}$ of a point in the elastic medium at rest. Moreover we use the metric signature $(+,-,-,-)$. As in general relativity, a general covariant formulation uses the metric associated with $\mathcal{R}$, written as $g_{\m\, \mu\nu}$ in this paper ; but this metric is one more time related to the considered elastic medium (as for instance the Poincaré-Lorentz transformation which is expressed with respect to $c_\m$, the propagation speed of the wave in the elastic medium).

The general covariant formulation (see {\it e.g.}~\cite{landau_th_ch}) of the Lagrangian density~(\ref{econc_l}), in a reference frame $\mathcal{R}$ with an elastic-medium metric tensor $g_{\m\, \mu\nu}$, is written as $\sqrt{-g_\m}\:\mathcal{L}$, where
\be \label{eacc_l}
\mathcal{L} = \frac{1}{2}\, \mathcal{T}\,\left(1\,+ \, \frac{\rho_0(\vec{r},t)}{\mu_0} \right) \left[ g_\m^{\,\mu\nu} \, \partial_\mu \varphi \, \partial_\nu \varphi \, - \, \frac{\Omega_\m^2}{c_\m^2} \varphi^2 \right] 
\ee
and $\sqrt{-g_\m}$ denotes the square root of the negative of the determinant of $g_{\m\, \mu\nu}$. (The action reads $S=\int \mathcal{L}\,\sqrt{-g_\m}\;\dd t\,\dd x^1\,\dd x^2\,\dd x^3$.)

\subsection{Wave equation and symbiosis equation}
The wave equation stems from the principle of least action (see \ref{sa_equas} for more details) and is written as
\be \label{eacc_ch}
\overset{\smallfrown}{\Box}_\m \varphi +  \frac{\Omega_\m^2}{c_\m^2}\,\varphi 
 = -\frac{\rho_0}{\mu_0}\,\left(\overset{\smallfrown}{\Box}_\m \varphi +  \frac{\Omega_\m^2}{c_\m^2}\,\varphi\,\right) - \frac{1}{\mu_0}\, g_\m^{\,\mu\nu} \,\partial_\mu(\rho_0)\,\partial_\nu\varphi \,.
\ee
where $\overset{\smallfrown}{\Box}_\m$ explicitly denotes the d'Alembert operator (specific to the elastic medium) in curvilinear coordinates, {\it i.e.} $ \overset{\smallfrown}{\Box}_\m \varphi = \frac{1}{\sqrt{-g_\m}}\,\partial_\mu\left(\sqrt{-g_\m} \: g_\m^{\,\mu\nu} \,\partial_\nu\varphi \right)$. (In this paper we are not interested in the equation of motion for the concretion because, as seen in~\cite{conc} and in the low-velocity approximation, this equation could very weakly perturb the velocity and trajectory of the concretion given by the guidance formula, {\it i.e.} by the wave equation.)

The concretion does not generate waves any longer when the two following conditions are fulfilled: ($i$) the wave is governed by the Klein-Gordon-like equation in curvilinear coordinates ($\overset{\smallfrown}{\Box}_\m \varphi +  \frac{\Omega_\m^2}{c_\m^2}\,\varphi = 0$) and ($ii$) the {\it symbiosis equation},
\be \label{eacc_symb}
g_\m^{\,\mu\nu} \,\partial_\mu(\rho_0)\,\partial_\nu\varphi = 0 \,,
\ee
is satisfied. As expected, the corresponding previous result~(\ref{e1_symb}) is generalized.

\subsection{Newtonian elastic-medium metric, constant in time}
In general relativity the simplest metric tensor used to describe a motion of a particle in a gravitational field under the low-velocity and small gravitational field approximation is a Newtonian metric ({\it e.g.}~\cite{landau_th_ch} \S 87). In our context, from now on we only study the case in which the elastic-medium metric, $g_{\m\, \mu\nu}$, is Newtonian and constant in time. $g_{\m\, \mu\nu}$ is then written as
\be \label{eacc_gmunu}
g_{\m\, \mu\nu} = \begin{pmatrix}g_{\m\, 00} & 0 & 0 & 0 \\ 0 & -1 & 0 & 0 \\ 0 & 0 & -1 & 0 \\ 0 &0 & 0 & -1 \\ \end{pmatrix} \,,
\ee
in which $g_{\m\, 00}$ is constant in time and conveniently written as 
\be \label{eacc_g00}
\sqrt{g_{\m\, 00}}=1+\epsilon(\vec{r}) \,.
\ee

In general relativity, $\epsilon(\vec{r})$ of a Newtonian metric constant in time is equal to the gravitational potential energy per unit mass and divided by $c^2$ (cf. {\it e.g.} \cite{landau_th_ch} \S 87), {\it i.e.} $m_0 \,c^2 \, \epsilon(\vec{r})$ is equal to the gravitational potential acting on a point mass $m_0$ at position $\vec{r}$. Similarly we define the potential energy acting on the concretion (if it was located at point $\vec{r}$) as
\be \label{eacc_V}
V(\vec{r}) = m_0\,c_\m^2\;\epsilon(\vec{r}) \,.
\ee

Finally, $\rho_0$, the mass density of the concretion expressed in its proper reference frame, is written (cf. {\it e.g.} \cite{landau_th_ch} \S 90) in $\mathcal{R}$ as
\be \label{eacc_rho0}
\rho_0 = m_0\,\sqrt{1-\mathcal{V}^2/c_\m^2}~\: \delta(\vec{r}-\vec{\xi}(t)) \,,
\ee
where $\mathcal{V}$ denotes the velocity of the concretion measured in terms of the proper time, that is, by an observer located at the given point (cf. {\it e.g.} \cite{landau_th_ch} \S 88) -- which implies that $\vec{\mathcal{V}}=\frac{1}{\sqrt{g_{\m\, 00}}}\,\vec{v}$, where the velocity of the concretion measured in $\mathcal{R}$ is $\vec{v}=\frac{\dd \vec{\xi}}{\dd t}$.

In the following, the case $|\epsilon| \ll 1$ corresponds to the small-potential approximation.

\subsection{Effective Schrödinger equation with an external potential \label{ssac_schro}}
In the Newtonian elastic-medium metric under the low-velocity and small-velocity approximation, by using the modulating wave $\psi$~(\ref{epsi_def}), the wave equation~(\ref{eacc_ch}) without source leads to $-2\,\ii\,\frac{\Omega_\m}{c_\m^2}\,\frac{\partial \psi}{\partial t} \,-\, \Delta \psi - \vec{\nabla}\epsilon \cdot \vec{\nabla}\psi + 2\,\epsilon\,\frac{\Omega_\m^2}{c_\m^2}\, \psi = 0$. (Calculations are very similar to the one detailed in the appendix of~\cite{conc}, in which the definition of the modulating wave $\psi$~(\ref{epsi_def}) is used and the term $\frac{\partial^2 \psi}{\partial t^2}$ is neglected in the low-velocity approximation as well as a term with $\epsilon\, \frac{\partial \psi}{\partial t}$.) When $\epsilon(\vec{r})$, as $\omega_\m(\vec{r})$ in the previous section, is very smooth comparatively to the wave ({\it i.e.} when the variation of $\epsilon(\vec{r})$ is very small over a wavelength of $\psi$), the term $\vec{\nabla}\epsilon \cdot \vec{\nabla}\psi$ is negligible. This leads to
\be \label{eacc_schro0}
\ii\,\frac{\partial \psi}{\partial t} = -\,\frac{c_\m^2}{2\,\Omega_\m}\Delta\psi \:+\: \epsilon\,\Omega_\m\,\psi \,.
\ee
It is worth noting here that this equation is a generalization of Eq.~(\ref{e1_schro0}) and is
obtained by similar techniques (in agreement with de Broglie's original ideas concerning the link between Klein-Gordon and Schrodinger equations, as seen {\it e.g.} in~\cite{ldb_tentative} \S \cRM{2}.7 ).

By using $\hbe$~(\ref{ehbe}) and the potential energy $V$~(\ref{eacc_V}) related to the concretion in $\mathcal{R}$, we get
\be \label{eacc_schro}
\ii\,\hbe\,\frac{\partial \psi}{\partial t} = -\,\frac{\hbe^2}{2\,m_0}\,\Delta\psi \,+\, V\,\psi \,.
\ee
This equation has the exact form as the Schrödinger equation with an external potential $V$.

\subsection{Assumption about the pulsation of the medium and the elastic-medium metric \label{ssge_major}} 
Let us now make a link between the elastic-medium metric, $g_{\m\, \mu\nu}$, and inhomogeneous PM, $\Omega_\m'(\vec{r})$. In general relativity, the metric $g_{\mu\nu}$ can be directly related to the gravitational field in a reference frame. It is then natural to assume in our toy system that the elastic-medium metric, $g_{\m\, \mu\nu}$, is also related to properties in the elastic medium. We have seen in Section~\ref{s1} that not only PM is related to a crucial property of the elastic medium, but also (even if there is some misleading in this formalization) can generate an external field applied on the concretion and on the transverse wave. By comparison with effective Schrödinger equations (Eqs.~(\ref{e1_schro0}) and (\ref{eacc_schro0})) it is very tempting to assume that $\omega_\m(\vec{r})$ is related to $\epsilon(\vec{r})$ in the Newtonian elastic-medium metric such that
\be \label{ege_equiv}
\epsilon(\vec{r})\: = \: \frac{\omega_\m(\vec{r})}{\Omega_\m}   \,.
\ee
This expression means that the potential energy $V(\vec{r})=\hbe\;\omega_\m(\vec{r})$ written in previous section~(\ref{e1_V}) is the same as the one in this section~(\ref{eacc_V}).

Then, we make the following assumption in our macroscopic toy system (under the low-velocity and small-potential approximation): in a system with the PM $\Omega_\m'(\vec{r})$ the elastic-medium metric $g_{\m\, \mu\nu}$ is Newtonian-like such that
\be \label{ege_g00}
\sqrt{g_{\m\, 00}}=\frac{\Omega_\m'(\vec{r})}{\Omega_\m} = 1 + \frac{\omega_\m(\vec{r})}{\Omega_\m} \,.
\ee
(Hence, for a system with uniform and reference PM, $\Omega_\m'(\vec{r})=\Omega_\m$, the elastic-medium metric is Minkowski-like.)

\subsection{From the symbiosis equation to the concretion guidance formula \label{ssac_guid}}
The symbiosis equation associated with the conservation of the mass $m_0$ of the concretion lead to a covariant guidance formula -- which yields the effective de Broglie-Bohm guidance formula in low-velocity approximation~\cite{conc}. By using Eq.~(\ref{eacc_symb}) and the mass continuity equation (see \ref{sa_guid} for more details) we get
\be \label{eacc_ur_guid}
\frac{1}{c_\m^2\,g_{\m\, 00}}\,\frac{\sqrt{1-\mathcal{V}^2/c_\m^2}}{\sqrt{g_{\m\, 00}}}\,\frac{\partial \varphi}{\partial t}\,\left[\frac{\dd }{\dd t}\frac{\sqrt{g_{\m\, 00}}}{\sqrt{1-\mathcal{V}^2/c_\m^2}} 
\right]\rho_0 \;+\; \left[\frac{\vec{v}}{c_\m^2\,g_{\m\, 00}}\,\frac{\partial \varphi}{\partial t} +  \vec{\nabla}\varphi\right]\cdot \vec{\nabla}\rho_0 \; = 0 \,,
\ee
(Recall that this relation only concerns a concretion studied in a referential $\mathcal{R}$
with a Newtonian elastic-medium metric~(\ref{eacc_gmunu}) constant in time.)

It is worth noting that $\rho_0$ (and then the Dirac delta function) must be seen as the limit of a derivable function (for instance a 3D Gaussian function). Recall that this is fully in agreement with the fact that our toy model could be interpreted as a limiting case of a wave monistic model, still to write (as discussed in Section \ref{ss1_syst}). Hence, when $\vec{\nabla}\rho_0=\vec{0}$ ({\it i.e.} at the maximum of $\rho_0$), the first term of the above equation is zero. This again implies that the second term is zero. So, Eq.~(\ref{eacc_ur_guid}) encapsulates the two following results:
\begin{eqnarray}
 &  & \frac{\dd }{\dd t}\frac{\sqrt{g_{\m\, 00}}}{\sqrt{1-\mathcal{V}^2/c_\m^2}} = 0 \,,  \label{eacc_guid_E} \\
 &  & \frac{\vec{v}}{c_\m^2\,g_{\m\, 00}}\,\frac{\partial \varphi}{\partial t}(\vec{\xi},t) +  \vec{\nabla}\varphi(\vec{\xi},t) = \vec{0} \label{eacc_guid} \,,
\end{eqnarray}
in which $g_{\m\, 00}$ is evaluated at the position of the concretion, $\vec{r}=\vec{\xi}$. These equations generalize the corresponding ones in Section~\ref{ss1_contradiction}.

Eq.~(\ref{eacc_guid_E}) has a direct counterpart in general relativity: the energy of a point mass $m_0$ in a static gravitational field ($\frac{\sqrt{g_{00}}\,m_0\,c^2}{\sqrt{1-\mathcal{V}^2/c^2}}$, cf. {\it e.g.} \cite{landau_th_ch} \S 88) is constant in time. Eq.~(\ref{eacc_guid}) yields (see below, Eq.~(\ref{eacc_vguid})), in the low-velocity approximation, an analogous de Broglie-Bohm guidance formula in quantum mechanics. (This equation, which concerns the transverse displacement wave $\varphi$ and not only its phase, is called in the following the $\varphi$-guidance formula.) It is very interesting to note that the symbiosis equation not only leads to a guidance formula, but also leads to an energy conservation. The symbiosis equation then appears to be very crucial.

So, when the concretion is in symbiosis with the wave (if this exists): ($i$) the transverse wave $\varphi$ is governed by a Klein-Gordon-like equation~(\ref{eacc_ch}) (without source of waves) and ($ii$) the motion of the concretion is given by the symbiosis equation~(\ref{eacc_symb}). Furthermore it is worth noting that the latter equation encapsulates both an energy conservation of the concretion and its guidance formula. (This means that when one of the previous results is not satisfied, for instance when the energy conservation of the concretion is not satisfied, the concretion and the wave are not in symbiosis.)  Finally, note that Eqs.~(\ref{eacc_guid_E}), (\ref{eacc_guid}) and the linear wave field~(\ref{eacc_ch}) do not depend on the maximum amplitude of the wave field $\varphi$.

On the contrary of the previous Section~\ref{ss1_contradiction} and by using a general covariant formulation, the problem concerning the guidance formula and the velocity of the concretion under the influence of a potential does not exist any more. Here, the velocity of the concretion under the influence of potential (\ref{e1_V}) or (\ref{eacc_V}) can change in time and this occurs in agreement with energy conservation~(\ref{eacc_guid_E}).

\subsubsection{In the low-velocity and small-potential approximation \label{ssac_guid_approx}}
In the low-velocity and small-potential approximation and by using the modulating wave $\psi$~(\ref{epsi_def}), the $\varphi$-guidance formula (\ref{eacc_guid}) yields $\vec{v}\,\psi(\vec{\xi},t)=\frac{c_\m^2}{\ii\:\Omega_\m}\,\vec{\nabla}\psi(\vec{\xi},t)$. It is then convenient to write $\psi$ as
\be \label{epsi_FPhi}
\psi = F\:\e^{\ii  \Phi} \,,
\ee
where the magnitude $F$ and the phase $\Phi$ are two real functions. Thus, the velocity of the concretion reads
\be \label{eacc_vguid}
\vec{v} = \frac{c_\m^2}{\Omega_\m} \: \vec{\nabla}\Phi(\vec{\xi},t) \,.
\ee 
This equation has a direct counterpart in quantum mechanics: the de Broglie-Bohm guidance formula~\cite{ldb_ondeguidee1927,bohm1} ({\it i.e.} by using notation~(\ref{epsi_FPhi}) and $\hbe$~(\ref{ehbe}),  $\vec{v}=\frac{\hbar}{m_0}\vec{\nabla}\Phi$)~\footnote{Nevertheless relation~(\ref{eacc_vguid}) is only valid at the location of the concretion. This corresponds to the de Broglie's {\it double solution} viewpoint. In the context of walkers, see~\cite{yc_wavepartduality11} for an acute discussion between the latter viewpoint and the one in bohmian mechanics.}.

There is another manner to express $\vec{v}$. The effective guidance formula~(\ref{eacc_vguid}) and the effective Schrödinger equation~(\ref{eacc_schro0}) (both based on the symbiosis between the transverse wave $\varphi$ and the concretion) yield
\bea \label{eacc_Q}
& & m_0\,\frac{\dd \vec{v}}{\dd t}=-\vec{\nabla}\left[Q(\vec{\xi},t) + V(\vec{\xi},t)\right] \,, \\ 
& & \mathrm{where}\hspace*{0.5cm} Q = -\,\frac{\hbe^2}{2\,\m_0} \; \frac{\Delta F}{F} \,.
\eea 
(Calculations are very similar to the ones in Bohmian mechanics, {\it e.g.}~\cite{applied_bohmian}, or in our context in the appendix of~\cite{conc}.) The wave potential $Q$ has the same form as the de Broglie-Bohm quantum potential~\cite{bohm1} (and also {\it e.g.}~\cite{ldb_tentative} \S \cRM{10}).

When the velocity of the concretion obeys the guidance formula, the amplitude of the transverse wave at the location of the concretion is such that $\vec{\nabla}{F}(\vec{\xi},t)=\vec{0}$. (Indeed, in the low-velocity and small potential approximation and by using the modulating wave $\psi$, Eq.~(\ref{eacc_guid}) becomes $\frac{-\ii\,\Omega_\m}{c_\m^2}\,\vec{v}\,\psi(\vec{\xi},t) + \vec{\nabla}\psi(\vec{\xi},t) = \vec{0}$, whose real part yields $\vec{\nabla}{F}(\vec{\xi},t)=\vec{0}$.) The concretion is then located at a local extremum of the vibration amplitude field $F(\vec{r},t)$. To give an image, apart from transverse oscillations, the concretion moves as if it surfs on the wave.

\subsection{Energetic considerations }
The energy $\Wconc$ and momentum $\vpconc$ of the concretion are defined naturally from time-averaged values (over one transverse oscillation) of the energy and momentum densities in the oscillating elastic medium around the location of the concretion. These energy and momentum densities are associated with the stress-energy tensor of the system, $T_{\alpha\beta}$, which results from the Lagrangian density of the system (cf. {\it e.g.} \cite{landau_th_ch} \S 94). Here (see \ref{sa_Econc} for more details) we get
\be \label{eacc_T}
T_{0\alpha} = \mathcal{T}\,\left(1\,+ \, \frac{\rho_0}{\mu_0} \right)
\left[\partial_0 \varphi \, \partial_\alpha \varphi - \frac{1}{2} g_{\m\, 0\alpha}\left(\partial_\mu \varphi \, \partial_\nu \varphi \, g_\m^{\mu\nu} - \frac{\Omega_\m^2}{c_\m^2} \varphi^2 \right)\right],
\ee
where $\rho_0$ is given by Eq.~(\ref{eacc_rho0}). Without $\rho_0$, {\it i.e.} without the concretion, this result is in agreement with the stress-energy tensor of the Klein-Gordon field in a metric $g_{\mu\nu}$ (cf. {\it e.g.} \cite{MBlau_LecturesNotes} \S 6.7), as expected.

When the concretion is assumed to be in symbiosis with the wave ({\it i.e.} the generalized $\varphi$-guidance formula~(\ref{eacc_guid}) is satisfied), expression~(\ref{eacc_T}) leads to (see \ref{sa_Econc} for more details) the energy $\Wconc$ and momentum $\vpconc$ of the concretion:
\bea \label{eacc_ur_EPconc}
& & \Wconc = \frac{1}{2}\,m_0\, \frac{\sqrt{1-\frac{\mathcal{V}^2}{c_\m^2}}}{\sqrt{g_{\m\, 00}}}
\cdot \left[\left(1+\frac{\mathcal{V}^2}{c_\m^2}\right)\, \left\langle\, \left(\frac{\partial \varphi(\vec{\xi},t)}{\partial t}\right)^2\; \right\rangle + g_{\m\, 00}\,\Omega_\m^2\, \langle\, \varphi^2(\vec{\xi},t)\, \rangle \right] \nonumber\\
& & \vpconc = m_0\, \frac{\sqrt{1-\frac{\mathcal{V}^2}{c_\m^2}}}{\sqrt{g_{\m\, 00}}} \: \left\langle\,\left(\frac{\partial \varphi(\vec{\xi},t)}{\partial t}\right)^2\; \right\rangle \: \frac{1}{g_{\m\, 00}\:c_\m^2}\: \vec{v}\,.
\eea 

Now, we have to distinguish two different cases. In the first one, $\Wconc$ and $\vpconc$ have a same form as in general relativity in a Newtonian metric constant in time. This case occurs when the concretion has a specific transverse oscillation which depends on the PM; the concretion is then called in its {\it reference state} (precisely defined in the following paragraph). The study of a concretion in an effective free fall, below in Section~\ref{ssge_fall}, corresponds to this case. The second case, more general than the first one but restricted to the small velocity and small potential approximation, provides for $\Wconc$ and $\vpconc$ a same form as in quantum mechanics. The study of a concretion in an effective harmonic potential, below in Section~\ref{ssge_pot}, corresponds to this second case.

\subsubsection{Concretion in its reference state and counterpart in general relativity \label{ssac_RefState}}
Following~\cite{conc} we define the reference state of the concretion when ($i$) the concretion is in symbiosis with the wave and when ($ii$) the relation
\be \label{eacc_etat_ref}
\left\langle \left(g_\m^{\, \mu\nu} \, \partial_\mu \varphi \, \partial_\nu \varphi \,  - \, \frac{\Omega_\m^2}{c_\m^2} \varphi^2\right)_{\vec{\xi}(t)}\: \right\rangle = 0 \,,
\ee
is satisfied, in which $\langle (\cdots)_{\vec{\xi}(t)} \rangle$ denotes the time-averaged value over one transverse oscillation at the location of the concretion. (We assume throughout this article that the time period of transverse oscillations is much shorter than the characteristic evolution time of the motion of the concretion given by $\vec{\xi}(t)$. This assumption is experimentally realized for walking droplets experiments and also for the bead sliding on a vibrating string~\cite{yc_sao1999}.) Taking into account the generalized $\varphi$-guidance formula~(\ref{eacc_guid}), Eq.~(\ref{eacc_etat_ref}) leads to $\left\langle\,\left(\frac{\partial \varphi(\vec{\xi},t)}{\partial t}\right)^2\; \right\rangle = \frac{g_{\m\, 00}}{1-\mathcal{V}^2/c_\m^2}\,\Omega_\m^2\, \langle\varphi^2(\vec{\xi},t)\rangle$. This implies that a concretion in its reference state and at rest in a referential $\mathcal{R}$ oscillates at the pulsation $\sqrt{g_{\m\, 00}}\:\Omega_\m$ in $\mathcal{R}$ -- which means, in a Newtonian-like metric, that the concretion oscillates at pulsation $\Omega_\m$ measured in its proper time. Taking into account relation~(\ref{ege_g00}) between $g_{\m\, 00}$ and PM, this simply means that the concretion has a transverse oscillation at pulsation equal to the PM at its location, $\Omega_\m'(\vec{\xi})$. 

It is also worth noting that the wave potential at the location of the concretion, $Q(\vec{\xi},t)$, is related to the reference state of the concretion in the low-velocity and small-potential approximation: when $Q(\vec{\xi},t)=0$ the concretion is in its reference state (this relation is established at the next subsection). This could be interesting because in quantum mechanics with the Bohmian point of view, the quantum potential $Q$ is responsible for deviations of Bohmian trajectories from classical behavior in classical mechanics ({\it e.g.}~\cite{applied_bohmian}).

Eqs.~(\ref{eacc_ur_EPconc}) and (\ref{eacc_etat_ref}) lead to
\bea \label{eacc_EPconc_ref}
& & \Wconc = m_0\, \frac{\sqrt{g_{\m\, 00}}}{\sqrt{1-\mathcal{V}^2/c_\m^2}} ~ \Omega_\m^2\, \langle\varphi^2(\vec{\xi},t)\rangle  \\
& & \vpconc = m_0 \,\frac{\vec{v} }{\sqrt{g_{\m\, 00}}\:\sqrt{1-\mathcal{V}^2/c_\m^2}} ~ \frac{\Omega_\m^2\, \langle\varphi^2(\vec{\xi},t)\rangle}{c_\m^2} .
\eea
Whenever the condition
\be \label{eacc_cond_reine}
\Omega_\m^2\, \langle\varphi^2(\vec{\xi},t)\rangle = c_\m^2 
\ee
is fulfilled, which means that the average quadratic transverse oscillation velocity of the
concretion is equal to $c_\m^2$, the energy and momentum of the concretion are written in the same form as in general relativity for a point mass $m_0$ in a static gravitational field
and in a Newtonian metric ({\it i.e.} $\frac{\sqrt{g_{00}}\,m_0\,c^2}{\sqrt{1-\mathcal{V}^2/c^2}}$ and $\frac{m_0\, \vec{v}}{\sqrt{g_{00}}\:\sqrt{1-\mathcal{V}^2/c^2}}$ respectively, {\it e.g.} \cite{landau_th_ch} \S 88 or \cite{ldb_ondes_mvt} \S \cRm{2}.2).

Finally, it is interesting to note that condition~(\ref{eacc_cond_reine}) means in particular that the amplitude of the transverse wave at the location of the concretion remains constant in time, {\it i.e.} $\frac{\partial F}{\partial t}(\vec{\xi},t)=0$. This condition associated with Eq.~(\ref{eacc_guid_E}) (which is deduced from the symbiosis equation) implies that the energy of the concretion remains constant in time.

\subsubsection{General case under the low-velocity and small-potential approximation and counterpart in quantum mechanics} 
Now, we consider the general case in which the concretion is not necessarily in its reference state but under the low-velocity and small-potential approximation. It is then convenient to write the pulsation $\omega$ and the wave vector $\vec{k}$ in the phase $\Phi$ of $\psi$~(\ref{epsi_FPhi}) as
\bea \label{epsi_w_k}
& & \omega(\vec{r},t) = -\,\frac{\partial\, \Phi(\vec{r},t)}{\partial\, t}\\
& & \vec{k}(\vec{r},t) = \vec{\nabla}\Phi(\vec{r},t) \,.
\eea
By using the pulsation $\omega$ at the location of the concretion and provided that condition~(\ref{eacc_cond_reine}) is fulfilled, the energy and momentum of the concretion in Eq.~(\ref{eacc_ur_EPconc}) are now written as
\begin{eqnarray}
& & \Wconc = m_0\, c_\m^2 \:+\: \hbe\;\omega(\vec{\xi},t)   \label{eacc_Econc} \\
& & \vpconc = m_0\: \vec{v} \label{eacc_Pconc} \,,
\end{eqnarray}
(see \ref{sa_Econc} for more details). The total energy of the concretion is equal to the effective rest mass energy of the concretion ($m_0\,c_\m^2$, as in relativity) plus an additional energy $\Econc=\hbe\;\omega(\vec{\xi},t)$ (as in quantum mechanics). It is worth noting that both the effective rest mass energy ($m_0\,c_\m^2$) and the coefficient $\hbe$ come naturally from the only condition~(\ref{eacc_cond_reine}).

It is particularly interesting to write $\Econc$ and $\vpconc$ by using $\psi$, the modulating wave. It is easy to show that Eq.~(\ref{eacc_Econc}) reads
\bea \label{eacc_Econc_psi}
 & & \Econc = -\,\hbe \: \frac{\partial\, \Phi(\vec{\xi},t)}{\partial\, t} \\
 & & \ii \,\hbe\: \frac{\partial \psi}{\partial t}(\vec{\xi},t)\, = \, \Econc \:\psi(\vec{\xi},t) \,,
\eea
and Eq.~(\ref{eacc_Pconc}),
\bea \label{eacc_Pconc_psi}
 & & \vpconc = \hbe\,\vec{\nabla}\Phi(\vec{\xi},t) \\ 
 & & \frac{\hbe}{\ii}\; \vec{\nabla}\psi(\vec{\xi},t)\, = \, \vpconc\: \psi(\vec{\xi},t) \,,
\eea
(where we have used relations~(\ref{ehbe}), (\ref{epsi_w_k}), $\frac{\partial F}{\partial t}(\vec{\xi},t)=0$ and $\vec{\nabla}{F}(\vec{\xi},t)=\vec{0}$). The additional energy $\Econc$ and the momentum $\vpconc$ of the concretion are then in exact agreement with the energy and momentum of an analogous quantum system -- apart from they specifically concern $\psi$ at the location of the concretion. As has been suggested by Louis de Broglie in quantum mechanics (see {\it e.g.} \cite{ldb_tentative}, \S \cRm{11}), in our macroscopic toy system the `particle' concretion accounts for quantities (here energy and momentum) commonly attributed to the wave-like nature of the system.

Finally, similarly to quantum mechanics with a particle point of view (see {\it e.g.} \cite{Q_th_motion}), the energy of the concretion is also written as  
\be \label{eacc_Econc_Q}
\Econc = \frac{1}{2}\,m_0\,v^2 + Q(\vec{\xi},t) + V(\vec{\xi},t) \,.
\ee
(Calculations are very common in Bohmian mechanics, {\it e.g.}~\cite{applied_bohmian}, and use in our context the real part of the effective Schrödinger equation with $\Psi=F\:\e^{\ii  \Phi}$.) Moreover it is easy to show (by using Eqs.~(\ref{epsi_def}), (\ref{eacc_Econc_psi}), (\ref{eacc_Pconc_psi}) and condition~(\ref{eacc_cond_reine})) that Eq.~(\ref{eacc_Econc_Q}) leads to
\be
\frac{1}{2}\,\left\langle \left(g_\m^{\, \mu\nu} \, \partial_\mu \varphi \, \partial_\nu \varphi \,  - \, \frac{\Omega_\m^2}{c_\m^2} \varphi^2\right)_{\vec{\xi}(t)}\: \right\rangle = \frac{Q(\vec{\xi},t)}{m_0\,c_\m^2} \,.
\ee
Therefore, the more $Q$, the more the concretion is out of its reference state.

\subsection{Effective free fall \label{ssge_fall}} 
\begin{figure}
\begin{minipage}[c]{0.5\linewidth}
\centering
\includegraphics[width=1 \columnwidth, clip=true]{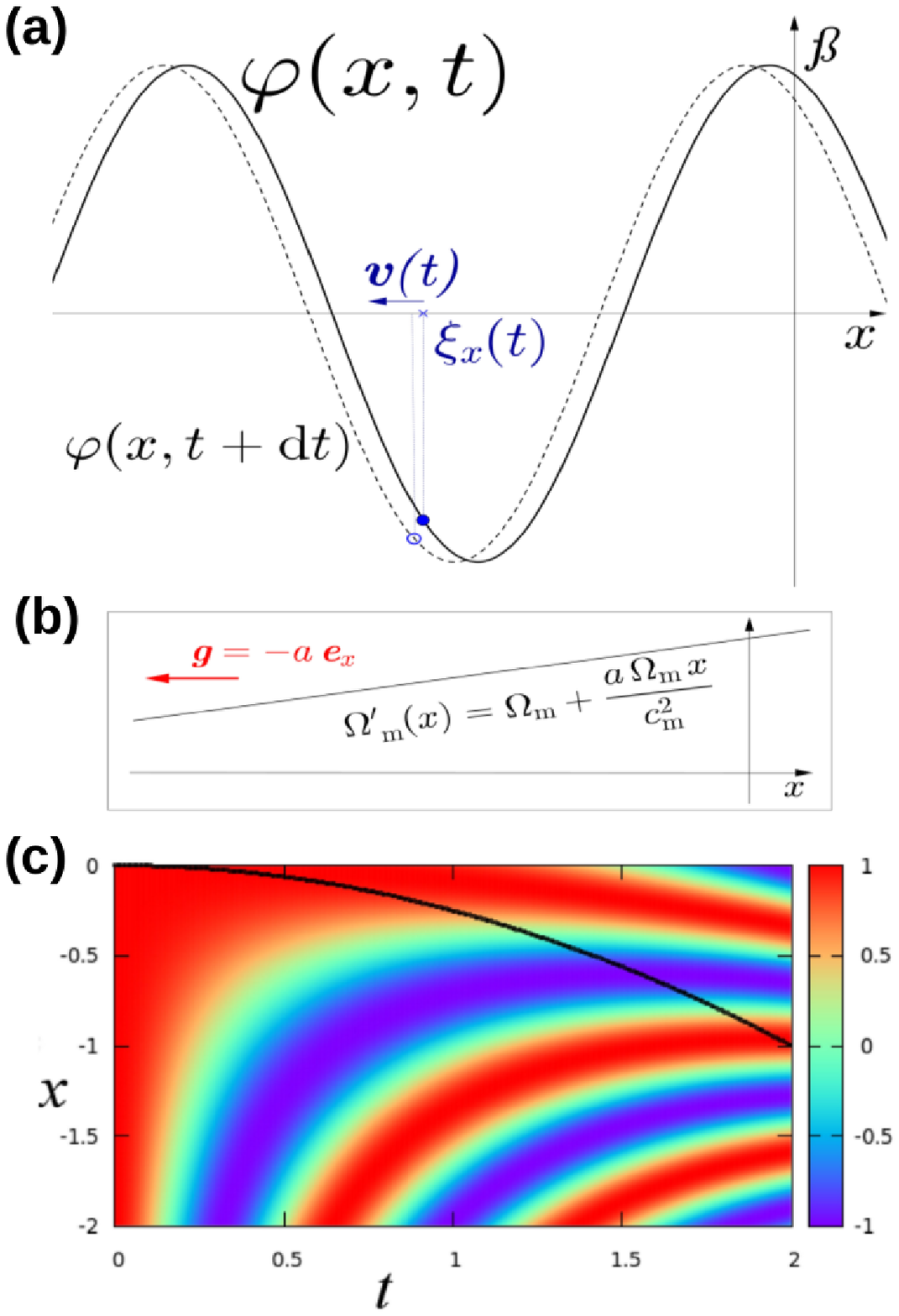}\hfill
\end{minipage}
\begin{minipage}[r]{0.3\linewidth}
\caption{Schematic representation of the system corresponding to an effective free fall. (a): Transverse wave $\varphi$ and the location of the concretion, $\xi$, at times $t$ and $t+\dd t$. (b): PM, $\Omega_\m'(\vec{r})$~(\ref{ege_omega_fall}), leading to the effective gravitational acceleration, $\vec{g} = - a\,\vec{e}_x$. (c): Stroboscopic view of the transverse wave $\varphi$~(\ref{ege_varphi_fall}) and the corresponding location of the concretion (black points). Each time $t$ is an integer multiple of $\frac{2\,\pi}{\Omega_\m}$. Parameters of the theoretical system: $c_\m=10~\mathrm{m \cdot s}^{-1}$, $\Omega_\m = 1000~\mathrm{s}^{-1}$ and $a=0.5~\mathrm{m \cdot s}^{-2}$. Color bar is expressed in $A$, the maximal amplitude of transverse oscillations; $x$ and $t$ are respectively expressed in m and s.}
\label{fig2}
\end{minipage}
\end{figure}

We now illustrate a system in which there is particular and simple potential-like energy $V$. The concretion is considered in symbiosis with the transverse wave. Let this potential correspond to the one of a free fall in a uniform effective gravitational acceleration $\vec{g}$. We here consider a (flat) 2D elastic medium $(Oxy)$. (Transverse oscillations are directed along the vertical axis.) Finally, calculations are performed under the low-velocity and small-potential approximation. (See Fig.~\ref{fig2} for a basic representation of the system.)

An effective potential $V(\vec{r})$~(\ref{e1_V}) generates an effective gravitational acceleration such that
\be \label{ege_g}
\vec{g}(\vec{r})=-\frac{c_\m^2}{\Omega_\m}\,\vec{\nabla}\omega_\m(\vec{r}) 
\ee
(because $\hbe = \frac{m_0\,c_\m^2}{\Omega_\m}$). When 
\be \label{ege_omega_fall}
\omega_\m(\vec{r})=\frac{a\,\Omega_\m\,x}{c_\m^2} \,,
\ee where $a$ is constant and has the dimension of an acceleration, the effective gravitational acceleration is uniform and equal to $\vec{g} = - a\,\vec{e}_x$, where $\vec{e}_x$ denotes the unit vector along the $(Ox)$ axis.

From now on, we only consider the solution~\cite{bohm_MQ} 
\be \label{ege_psi_fall}
\psi(\vec{r},t)\,=\,\e^{-\ii\,\Omega_\m (a\,x\,t/c_\m^2 \,+\, \frac{1}{6}\,a^2\,t^3/c_\m^2)} 
\ee
for the corresponding effective Schrödinger equation~(\ref{eacc_schro}) with $\omega_\m(\vec{r})=\frac{a\,\Omega_\m\,x}{c_\m^2}$. Definition~(\ref{epsi_def}) of the modulating wave $\psi$ leads to the (real-valued) transverse wave:
\be \label{ege_varphi_fall}
\varphi(\vec{r},t)\,=A\:\cos\left(\Omega_\m (1+\frac{a\,x}{c_\m^2})\,t \,+\, \frac{\Omega_\m\,a^2\,t^3}{6\,c_\m^2}\right) \,.
\ee

Let us now investigate the velocity $\vec{v}$ and the location $\vec{\xi}$ of the concretion. According to $\psi$ given by Eq.~(\ref{ege_psi_fall}), the phase~(\ref{epsi_FPhi}) is $\Phi(\vec{r},t)=-\,\Omega_\m (a\,x\,t/c_\m^2 \,+\, \frac{1}{6}\,a^2\,t^3/c_\m^2)$. The guidance formula~(\ref{eacc_vguid}) leads to $\vec{v}=-a\,t\,\vec{e}_x$. This implies (because $\vec{v}=\frac{\dd \,\vec{\xi}}{\dd \, t}$) that $\vec{\xi}(t)=-\,\frac{1}{2}\,a\,t^2\,\vec{e}_x$, where we have assumed that $\xi_y(t=0)=0$. It is interesting to note that neither waves $\phi$ or $\psi$ nor the velocity of the concretion do not depend on the mass $m_0$ of the concretion.

Let us now evaluate the additional energy of the concretion, $\Econc$, and the momentum $\vpconc$ of the concretion. According to Eqs.~(\ref{eacc_Pconc}) or (\ref{eacc_Pconc_psi}) (and provided that condition (\ref{eacc_cond_reine}) is fulfilled), $\vpconc = -\,m_0\,a\,t\,\vec{e}_x$. The pulsation~(\ref{epsi_w_k}), $\omega(\vec{r},t)=-\frac{\partial \Phi(\vec{r},t)}{\partial t}$, at the location of the concretion is $\omega(\vec{\xi},t)=0$ (because $\xi_x(t)=-\,\frac{1}{2}\,a\,t^2$). According to Eqs.~(\ref{eacc_Econc}) or (\ref{eacc_Econc_psi}), the additional energy of the concretion, $\Econc$, is then zero and remains constant in time. (The total energy of the concretion, $\Wconc$, remains then constant in time and equal to $m_0\,c_\m^2$.)

It then appears that both motion and energy of the concretion in symbiosis with the wave $\varphi$~(\ref{ege_varphi_fall}) are in agreement with corresponding motion and energy of a point mass in a uniform gravitation field $\vec{g}=- a\,\vec{e}_x$ in classical mechanics (in which the gravitation potential energy transfers in kinetic energy). In other words, the concretion behaves as a point mass in classical mechanics moving in a potential $V$~(\ref{e1_V}). This is fully in agreement with the fact that $Q(\vec{\xi})=0$ (and then the concretion is in its reference state).


\subsection{Concretion in an effective harmonic potential \label{ssge_pot}}
We now study a system with an effective linear harmonic potential $V(x) = \frac{1}{2}\,m_0\,\omega_\mathrm{osc}^2\,x^2$, which corresponds to a linear harmonic oscillator in quantum mechanics. We here consider a (flat) 1D elastic medium $(Ox)$ (where transverse oscillations are directed for instance along the vertical axis). The wave $\varphi$ is considered as standing in the laboratory reference frame $\mathcal{R}$ and, in addition, in symbiosis with the concretion. Calculations are performed under the low-velocity and small-potential approximation.

According to Eq.~(\ref{e1_V}) the effective potential $V(x)$ is generated by modifying PM such that
\be \label{ege_omega_pot}
\omega_\m(x)=\frac{\omega_\mathrm{osc}^2\,\Omega_\m\,x^2}{2\,c_\m^2} \,.
\ee 

A solution of the effective Schrödinger equation~(\ref{eacc_schro}) with this effective potential and for a standing modulating wave $\psi(x,t)=F(x)\:\e^{-\,\ii\,\omega\,t}$ is such that (cf. {\it e.g.}~\cite{landau_MQ} \S 23) $F(x) \propto \e^{-\frac{m_0\,\omega_\mathrm{osc}}{2\,\hbe}\,x^2}\;H_n\left(\sqrt{\frac{m_0\,\omega_\mathrm{osc}}{\hbe}}\:x\right)$ and $\omega = \omega_\mathrm{osc}\,\left(n+\frac{1}{2}\right)$, where $H_n$ denotes a Hermite polynomial of degree $n$ ($n$ being a natural number). Due to the effective guidance formula~(\ref{eacc_vguid}), the velocity of the concretion is zero. (This is in agreement with the study of a harmonic oscillator with the Bohmian point of view, {\it e.g.} \cite{Q_th_motion} \S 4.9.) Moreover the concretion is located at a local extremum of the vibration amplitude field $F(x)$ (cf. Section~\ref{ssac_guid_approx}), for instance $\xi = 0$ for $n=0$ and $\xi = \pm \sqrt{\frac{\hbe}{m_0\,\omega_\mathrm{osc}}}$ for $n=1$. According to Eqs.~(\ref{eacc_Econc}) and (\ref{eacc_Pconc}) (and provided that condition (\ref{eacc_cond_reine}) is fulfilled), the additional energy of the concretion is $\Econc = \hbe\, \omega_\mathrm{osc}\,\left(n+\frac{1}{2}\right)$ and its momentum $\vpconc = \vec{0}$. Here, the wave potential $Q$ (\ref{eacc_Q}) is not zero. By using the expression of $F$ and a little bit of algebra Eq.~(\ref{eacc_Q}) leads to $Q(\xi) = \frac{\hbe^2}{2\,\m_0} \,\left[\frac{2\,m_0\,\omega_\mathrm{osc}}{\hbe}\,(n+\frac{1}{2}) \:-\: \frac{m_0^2\,\omega_\mathrm{osc}^2}{\hbe^2}\,\xi^2\right]$. This is of course in agreement with expression~(\ref{eacc_Econc_Q}) of $\Econc$. Finally, it is worth noting that the concretion is not in its reference state (because $Q(\xi) \neq 0$) and here the (standing) transverse wave $\varphi$ oscillates at pulsation $\Omega_\m + \omega$ (cf. (Eq.~\ref{epsi_def})).

This simple example is particularly interesting because it sheds light on a surprising and peculiar form of energy of the concretion. The velocity of the concretion is here zero. Consequently, the kinetic energy in relativistic and in classical mechanics of a mass $m_0$ should also be zero. But here, the concretion has an additional energy ($\Econc=\hbe\,\omega$, like in quantum systems) to its effective rest mass energy ($m_0\,c_\m^2$). We can interpret this result as follows: The concretion is `more' than an usual point mass $m_0$ as it would have been in common classical and relativistic systems. The concretion (a point-like high elastic medium density) shares properties both of the elastic medium ({\it i.e.} the PM at its location) and of the transverse wave with which it is in symbiosis. Then, the fact that the concretion is here embedded in a standing transverse wave oscillating with a pulsation greater than $\Omega_\m$ leads to an energy of a concretion greater than $m_0\,c_\m^2$. In other words, the energy of the concretion can be split in two different origins: one from its motion along the elastic medium and the other one from its transverse vibration (the latter provides in particular an effective rest mass energy $m_0\,c_\m^2$). Thus, even if the concretion has no kinetic energy along the elastic medium, its total energy can be greater than $m_0\,c_\m^2$ coming from its transverse oscillation -- and its transverse kinetic energy. (We note that common effective relativistic expression for energy is naturally extended in our context, when the concretion is out of its reference state ({\it i.e.} $Q(\xi) \neq 0$).)

This peculiar form of energy of the concretion is reminiscent with the ``variable proper mass'', $M_0$, suggested by de Broglie -- which played a great role in his work, for example for the motion of a particle~\cite{ldb_interpretation1987,ldb_reinterpretation}. (By using notations in (\ref{epsi_def}) and (\ref{epsi_FPhi}) and $m_0$ the proper mass of the particle, $M_0 = \sqrt{m_0^2 + \frac{\hbar^2}{c^2}\frac{\Box F}{F}}$; which becomes in the low-velocity approximation, $M_0 = m_0 + Q/c^2$, where $Q$ is here the quantum potential.) But more importantly in the context of walking droplets (which are also `more' than a usual point mass in classical mechanics), since the vertical/transverse speed is approximately 10 times larger than the horizontal speed, the horizontal kinetic energy could be considered as a usual kinetic energy (as in classical mechanics or in non-relativistic quantum mechanics) while the vertical kinetic energy (approximately 100 times larger than the horizontal one) could be considered as an additional or internal energy, which plays the role of de Broglie's variable proper mass~\cite{bush_private-com}. (In this vein it is interesting to note that recent studies on walkers~\cite{tadrist_waves18,tadrist_interaction2Walkers,bush_ratchets18,bush-review_modeling18,bush_phase18} highlight the role of their vertical motion~\cite{blanchette_vertical_motion} on the wave and the coupling between a drop's vertical and horizontal motion, which plays in particular a crucial role in systems of two interacting walkers, in orbiting pairs~\cite{yc_JFM06,yc_orbits08,bush_orbiting17} as well as in promenading pairs~\cite{wave-mediated,bush_promenades18}~\footnote{If the proximity between our toy model and walking droplets systems still holds, one could expect that the vertical energy of walkers increases with the memory and also partially accounts for the binding energy of interacting walkers.}.)

\section*{Conclusion}
In this paper we study a simple system -- inspired by walking droplets experiments with non-uniform fluid depth in the container -- in which a non-uniform pulsation of the medium (PM) is present. (A PM at point $\vec{r}$ reads $\Omega_\m'(\vec{r})=\Omega_\m + \omega_\m(\vec{r})$, where $\Omega_\m$ is the reference pulsation in this paper, as in~\cite{conc}; $\Omega_\m'(\vec{r})$ corresponds to the pulsation of transverse standing oscillations that the elastic medium of the system tends to support at point $\vec{r}$.) Recall that a submerged barrier in the vibrating container generates a kind of potential barrier acting both on walking droplets and on surface waves.

We first show that a Lorentz-Poincaré covariant description of the system leads to inconsistencies and contradictions. (The velocity of the concretion in symbiosis with the wave would have been constant in time, whatever the external potential). The contradiction is solved by using a general covariant formulation. In this paper we use a Newtonian-like metric tensor, which corresponds in general relativity to a gravitational field under the small-velocity and small-potential approximation. (Recall that almost all our calculations are also performed under this approximation.) It is important to stress that the metric tensor of the elastic medium is related to non-uniform PM~(\ref{ege_g00}). So then, a system with non-uniform PM could generate any effective potential, $V(\vec{r})=\hbe\:\omega_\m(\vec{r})$, acting both on the `particle' concretion and on the transverse wave.

Results can be grouped in two parts. The first one generalizes, in presence of an effective potential, previous results seen in~\cite{conc}. We nevertheless would like to emphasize the three following points. ($i$) Here again, the energy and momentum of the concretion have the same form as in relativity (in a static gravitational field in a Newtonian metric) and in quantum mechanics. ($ii$) The symbiosis equation~(\ref{eacc_symb}) appears to be very crucial. This equation (combined with the mass conservation of the
concretion) not only leads to the general covariant guidance formula~(\ref{eacc_guid}) (which yields the effective de Broglie-Bohm guidance formula in the small-velocity and small-potential approximation) but also to Eq.~(\ref{eacc_guid_E}) (which implies the conservation of energy in presence of an external potential). ($iii$) Last but not least, the modulating wave $\psi$ (defined from the transverse wave~(\ref{epsi_def})) is governed by a strictly analogous Schrodinger equation, of course with an external potential $V$.

The second part of results concerns the gravitational-like nature of the effective potential $V$ and a basic connection with general relativity (in the context of our macroscopic toy system). $V$ results from non-uniform PM. In parallel, non-uniform PM implies that proper time elapses differently at different points of the elastic medium (as in a static gravitational field in general relativity). (Otherwise, the Poincaré-Lorentz covariant formalization of the system leads to inconsistency and contradiction.) This means that the effective potential $V$ is an effective gravitational-like potential. Furthermore, it is worth noting that a proper time elapsing differently at different points of space as well as the relation between the proper time and the `world time' are easy to interpret in our toy system. In general relativity and in a Newtonian metric constant in time, the gravitational field implies a relation between the proper time $\tau$ and the `world time' $t$ (more precisely $\dd \tau = \sqrt{g_{00}}\;\dd t$, {\it e.g.} \cite{landau_th_ch} \S 88) which is not easy to interpret at first glance (because this directly involves the gravitational field). On the contrary, in our toy system, the effective potential, $V(\vec{r})=m_0\,c_\m^2\frac{\omega_\m(\vec{r})}{\Omega_\m}$, stems from the PM $\Omega_\m'(\vec{r})=\Omega_\m \cdot (1+\frac{\omega_\m(\vec{r})}{\Omega_\m})$ at point $\vec{r}$. The pulsation of the medium at a given point plays the role of a natural clock at this point (by the period of time coming from $\Omega_\m'(\vec{r})$). Due to relation~(\ref{ege_g00}) between the metric tensor of the elastic medium and PM (and then the effective potential), the relation between the `world time' and the proper time at this point reads: $\Omega_\m \: \dd \tau = \Omega_\m'(\vec{r}) \: \dd t$. This is just a comparison between the `world time' given by the natural clock at point $\vec{r}$ and the reference clock (oscillating at the reference pulsation $\Omega_\m$).

It could be interesting to very briefly connect our toy model to de Broglie's idea about internal quantum clock. According to him, a particle could be likened to a small clock whose frequency ({\it i.e.} its internal vibration) depends on the ``variable proper mass'' ({\it e.g.}~\cite{ldb_interpretation1987}). In our context, this is the elastic medium itself which can be interpreted as having a small clock at each point ({\it i.e.} its PM $\Omega_\m'(\vec{r})$) rather than only the `particle' concretion. Moreover, when the concretion is in its reference state, the concretion (at rest in the referential frame) has a pulsation of internal vibration equal to the PM at its location. Incidentally, we also note that the guidance formula results in de Broglie's idea from ``harmony of phases'' (according to which the particle's internal vibration is constantly in phase with the wave on which it is carried~\cite{ldb_interpretation1987}) while our general covariant guidance formula results also from a kind of harmony (called the symbiosis state) between the wave and the concretion, in which the concretion is no longer the source of the wave.

Our toy model, inspired by walking droplets experiments, naturally combines some effective quantum equations with some effective relativistic concepts, even with a small effective gravitational field. (Walking droplets systems, well-known for some quantum-like phenomena, also exhibit a kind of analogy with relativistic mechanics, as for instance an effective speed-dependent mass of walkers~\cite{bush_mass14}.) Nevertheless, at this step, our toy model does not present any stochastic phenomena. In order to give it randomness, one maybe could be inspired by the de Broglie–Bohm–Vigier approach in quantum mechanics~\cite{chebotarev_introdBBV} and/or recent studies~\cite{durt_afldb18,durt_chaos18} in a context closer to walking droplets systems.


To conclude we first mention the fact that quantum-like phenomena as well as general relativity-like phenomena also occur in a classical system has recently been studied in a very interesting paper~\cite{fort_cl_unruh}, therein with water waves. Next, we note that a possible kind of effective gravity field generated by the concretion has not been investigated in this article. In other words, the concretion behaves as a test particle in an effective gravitational field. (Let us also mention another approach with quantum solitons~\cite{durt_1,durt_2} -- also inspired by walking droplets experiments and de Broglie's double solution program -- in which the appearance of an effective gravitation is predicted.) It could be interesting in forthcoming studies to investigate if and how the `particle' concretion could also generate an effective gravitation.

\section*{Acknowledgments}
We gratefully thank Thomas Durt who has largely contributed to improve this paper, Yves Couder for very insightful comments and constructive suggestions, and John bush for help and discussion. Thanks to Thibaut Barthélémy for his help with English.

\renewcommand{\theequation}{A\arabic{equation}}
\setcounter{equation}{0}  

\renewcommand{\thesection}{Appendix A}
\section{Calculation of the wave equation \label{sa_equas}}
The wave equation comes from a principle of least action, when the wave field is subjected to a small change, $\varphi \rightarrow \varphi + \delta\varphi$, while the 4-position of the concretion is not subjected to a small change.

$\varphi \rightarrow \varphi + \delta\varphi$ leads to $\delta(\varphi^2) = 2\varphi\,\delta\varphi$ and $\delta(g_\m^{\,\mu\nu} \, \partial_\mu \varphi \, \partial_\nu \varphi)=2\, g_\m^{\,\mu\nu} \, \partial_\nu \varphi \, \partial_\mu (\delta\varphi)$. According to Eq.~(\ref{eacc_l}) conditions for which the small change $\delta \varphi$ imply at first order $\delta (\int \mathcal{L}\,\sqrt{-g_\m}\: \dd t\, \dd^3 r) = 0$, where boundary values are fixed, are written as
\be 
\int \mathcal{T}\left(1+\frac{\rho_0(\vec{r},t)}{\mu_0}\right)\left[g_\m^{\,\mu\nu} \, \partial_\nu \varphi \, \partial_\mu (\delta\varphi) - \frac{\Omega_\m^2}{c_\m^2}\,\varphi\,\delta\varphi \right]
 \cdot \sqrt{-g_\m}\: \dd t\, \dd^3 r = 0 \,.
\ee
Integrating by parts the term with $\partial_\mu (\delta\varphi)$, with fixed end points, and since the small change $\delta\varphi$ is arbitrary, lead to the wave equation~(\ref{eacc_ch}). (Note that the generalized Euler-Lagrange equation in curvilinear coordinates, $\frac{\partial \mathcal{L}}{\partial \varphi}=\frac{1}{\sqrt{-g_\m}}\,\partial_\mu \, \frac{\partial \sqrt{-g_\m}\, \mathcal{L}}{\partial (\partial_\mu \varphi)}$, leads to the same result, as expected.)

\renewcommand{\thesection}{Appendix B}
\section{Combining the symbiosis equation with the mass continuity equation \label{sa_guid}}
The mass continuity equation for the concretion is written in the considered elastic medium as $\frac{1}{\sqrt{-g_\m}}\:\partial_\mu (\sqrt{-g_\m}\: \rho_0 \, U^\mu) = 0$, where $g_\m$ is the determinant of the metric tensor and $U^\mu=c_\m\,\frac{\dd \xi^\mu}{\dd s_\m}$ is the 4-velocity of the concretion. In the Newtonian elastic-medium metric, $-g_\m = g_{\m\, 00}$ and $\dd s_\m = \sqrt{g_{\m\, 00}}\,\sqrt{1-\frac{\mathcal{V}^2}{c_\m^2}}\:c_\m\,\dd t$. This leads to
\be 
\rho_0 \left(\frac{\partial}{\partial t}\frac{1}{\sqrt{1-\frac{\mathcal{V}^2}{c_\m^2}}} + v^i\,\partial_i \frac{1}{\sqrt{1-\frac{\mathcal{V}^2}{c_\m^2}}} + \frac{1}{\sqrt{1-\frac{\mathcal{V}^2}{c_\m^2}}} \partial_i v^i \right)  \; + \;\frac{1}{\sqrt{1-\frac{\mathcal{V}^2}{c_\m^2}}} \left(\frac{\partial}{\partial t}\rho_0 + v^i\,\partial_i \rho_0\right) =  0\,,
\ee
in which $v^i$ is the $i$-th component of the vector velocity of the concretion, $\vec{v}=\frac{\dd \vec{\xi}}{\dd t}$, expressed in the laboratory reference frame $\mathcal{R}$. It is convenient to use the vector gradient, $\vec{\nabla}$, whose $i$-th component is $\partial_i$, and the particle derivative $\frac{\dd }{\dd t}= \frac{\partial }{\partial t} + v^i\partial_i$. Moreover, in the Newtonian elastic-medium metric: $\partial_i v^i = \frac{\vec{v}}{\sqrt{g_{\m\, 00}}}\cdot\vec{\nabla}\sqrt{g_{\m\, 00}}$. This result is for instance obtained by considering a Galilean elastic-medium reference frame $\mathcal{R}_X$ ({\it i.e.} its metric is Minkoswski-like) which coincides with $\mathcal{R}$ at a given time. (Indeed in $\mathcal{R}_X$, with spatio-temporal coordinates $(\vec{R},T)$: $\dd T= \sqrt{g_{\m\, 00}}\: \dd t$, $\dd X^i = \dd x^i$, $\mathcal{V}^i = \frac{v^i}{\sqrt{g_{\m\, 00}}}$ and $\frac{\partial \mathcal{V}^i}{\partial X^i}=0$.) Hence, the continuity equation becomes
\be \label{esa_cont}
\rho_0 \left(\sqrt{1-\mathcal{V}^2/c_\m^2}\:\frac{\dd }{\dd t}\frac{1}{\sqrt{1-\frac{\mathcal{V}^2}{c_\m^2}}} +  \frac{\vec{v}}{\sqrt{g_{\m\, 00}}}\cdot\vec{\nabla}\sqrt{g_{\m\, 00}}\right) \; + \; \frac{\partial}{\partial t}\rho_0 \;+\; \vec{v}\cdot \vec{\nabla}\rho_0 = 0 \,.
\ee

The symbiosis equation~(\ref{eacc_symb}) becomes in the Newtonian elastic-medium metric:
\be 
\frac{1}{c_\m^2\,g_{\m\, 00}}\,\frac{\partial \rho_0}{\partial t}\,\frac{\partial \varphi}{\partial t} \;-\; \vec{\nabla}\rho_0 \cdot \vec{\nabla}\varphi \; = 0 \,.
\ee
By substituting the term $\frac{\partial \rho_0}{\partial t}$ in Eq.~(\ref{esa_cont}), we get
\be \label{esa_ur_guid}
\frac{\rho_0\,\frac{\partial \varphi}{\partial t}}{c_\m^2\,g_{\m\, 00}} \left[\sqrt{1-\mathcal{V}^2/c_\m^2}\:\frac{\dd }{\dd t}\frac{1}{\sqrt{1-\frac{\mathcal{V}^2}{c_\m^2}}} + \frac{\vec{v}}{\sqrt{g_{\m\, 00}}}\cdot\vec{\nabla}\sqrt{g_{\m\, 00}}\right] \; + \; \left[\frac{1}{c_\m^2\,g_{\m\, 00}}\,\frac{\partial \varphi}{\partial t}\,\vec{v} +  \vec{\nabla}\varphi\right]\cdot \vec{\nabla}\rho_0 = 0 \,.
\ee
Moreover, $\vec{v}\cdot\vec{\nabla}\sqrt{g_{\m\, 00}} = \frac{\dd }{\dd t}\sqrt{g_{\m\, 00}}$ because the elastic-medium metric is constant in time ({\it i.e.} $\frac{\partial}{\partial t} g_{\m\, 00} = 0$). Then, Eq.~(\ref{esa_ur_guid}) leads to Eq.~(\ref{eacc_ur_guid}) in the main text.

\renewcommand{\thesection}{Appendix C}
\section{Energy and linear momentum of the concretion \label{sa_Econc}}
The energy density and the momentum density are evaluated from the stress–energy tensor density, $\sqrt{-g_\m}\,T$ of the system. $T_{\alpha\beta} = \frac{2}{\sqrt{-g_\m}}\;\frac{\partial \sqrt{-g_\m}\mathcal{L}}{\partial g_\m^{\,\alpha\beta}} ~-~ \frac{2}{\sqrt{-g_\m}}\; \partial_\gamma\, \frac{\partial \sqrt{-g_\m}\mathcal{L}}{\partial (\partial_\gamma g_\m^{\,\alpha\beta})}$ (cf. {\it e.g.} \cite{landau_th_ch} \S 94). According to Eq.~(\ref{eacc_l}) and in the Newtonian elastic-medium metric (and using $\frac{\partial \sqrt{-g_\m}}{\partial g_\m^{\,\alpha\beta}} = -\frac{1}{2}\,\sqrt{-g_\m}\:g_{\m\, \alpha\beta}$, cf. {\it e.g.} \cite{landau_th_ch} \S 86) we get 
\be \label{esa_T}
T_{0\alpha} =  \mathcal{T}\,\left(1\,+ \, \frac{\rho_0}{\mu_0} \right)
 \cdot \left[\partial_0 \varphi \, \partial_\alpha \varphi \; - \frac{1}{2} g_{\m\, 0\alpha}\left(\partial_\mu \varphi \, \partial_\nu \varphi \, g_\m^{\mu\nu} \, - \, \frac{\Omega_\m^2}{c_\m^2} \varphi^2 \right)\right] \,.
\ee
\vspace*{0.5\baselineskip}

In general relativity, energy and the momentum of a particle moving in a Newtonian metric constant in time is given by its covariant 4-impulsion $p_\alpha$ (cf. {\it e.g.}~\cite{landau_th_ch} \S 88 and \cite{MBlau_LecturesNotes} \S 2). Furthermore, for a particle of mass $m_0$, $c \cdot p^\alpha$ (where $p^\alpha=m_0\,U^\alpha$ is its contravariant 4-impulsion) is equal to the spatial integration of its corresponding stress-energy tensor density, $\sqrt{-g}\:T^{0\alpha}$ (see {\it e.g.} \cite{JFranklin} \S 5.8 for a detailed calculation), {\it i.e.}. $\int \sqrt{-g}\:T^{0\alpha}\, \dd^3r = c\,p^\alpha$. (In this context, $T^{\alpha\beta}=\rho_0\,U^0\,U^\alpha$; verification of calculations is very easy in a Newtonian metric constant in time by using analogous expression~(\ref{eacc_rho0}) for $\rho_0$.) In our context (corresponding to a transverse oscillating concretion) the spatial integration around the concretion does not suffice: we also have to take the average value over one transverse oscillation (written as $\langle\cdots\rangle$). Hence, in a Newtonian elastic-medium metric and taking into account covariant versus contravariant expressions, the 4-impulsion $p_\alpha$ of the concretion (where $p_0  = \Wconc/c_\m$ and $p_i=-\vpconc$) is equal to the time-averaged value during one transverse oscillation of the spatial integration around the concretion of $\frac{1}{c_\m\,\sqrt{g_{\m\,00}}}\,T_{0\alpha}$. Hence, $\int \frac{\langle T_{00} \rangle}{\sqrt{g_{\m\, 00}}}\, \dd^3r = \Wconc$ and $\int \frac{\langle T_{0i} \rangle}{\sqrt{g_{\m\, 00}}}\, \dd^3r = c_\m \, p_i$, where the integration is made around the location of the concretion. Then, according to expression~(\ref{esa_T}) for the stress-tensor, Eq.~(\ref{eacc_rho0}) for $\rho_0$ and the fact that the concretion is in symbiosis with the wave ({\it i.e.} the generalized $\varphi$-guidance formula~(\ref{eacc_guid}) is satisfied), we get
\bea \label{esa_ur_EPconc}
& & \Wconc = \frac{1}{2}\,m_0\, \frac{\sqrt{1-\frac{\mathcal{V}^2}{c_\m^2}}}{\sqrt{g_{\m\, 00}}} \cdot \left[\left(1+\frac{\mathcal{V}^2}{c_\m^2}\right)\, \left\langle\, \left(\frac{\partial \varphi(\vec{\xi},t)}{\partial t}\right)^2\; \right\rangle + g_{\m\, 00}\,\Omega_\m^2\, \langle\, \varphi^2(\vec{\xi},t)\, \rangle \right]  \nonumber \\
& & \vpconc = m_0\, \frac{\sqrt{1-\frac{\mathcal{V}^2}{c_\m^2}}}{\sqrt{g_{\m\, 00}}} \: \left\langle\,\left(\frac{\partial \varphi(\vec{\xi},t)}{\partial t}\right)^2\; \right\rangle \: \frac{1}{g_{\m\, 00}\:c_\m^2}\: \vec{v}\,,
\eea 
in which $\vec{\mathcal{V}}=\frac{1}{\sqrt{g_{\m\, 00}}}\,\vec{v}$. (Recall that $\mathcal{V}$ denotes the velocity of the concretion measured in terms of the proper time.) \\

Let us now consider the general case in which the concretion is not necessarily in its reference state. However the calculation is carried out under the low-velocity and small-potential approximation. By using Eqs.~(\ref{epsi_def}), (\ref{epsi_FPhi}) (\ref{epsi_w_k}) and condition~(\ref{eacc_cond_reine}) we get $\left\langle\, \left(\frac{\partial \varphi(\vec{\xi},t)}{\partial t}\right)^2\; \right\rangle = \left(\Omega_\m + \omega(\vec{\xi},t)\right)^2\,\langle \varphi^2(\vec{\xi},t)\,\rangle$ -- because the magnitude $F$ at the location of the concretion is constant in time. In first-order approximation in $\frac{v^2}{c_\m^2}$, $\frac{\omega}{\Omega_\m}$ (cf.~\cite{conc} Appendix) and $\epsilon$, Eqs.~(\ref{esa_ur_EPconc}) become
\bea 
& & \Wconc = m_0\; \Omega_\m^2\, \langle\varphi^2(\vec{\xi},t)\rangle \; \left(1 + \frac{\omega(\vec{\xi},t)}{\Omega_\m} \right) \\
& & \vpconc = m_0\; \Omega_\m^2\, \langle\varphi^2(\vec{\xi},t)\rangle \; \frac{\vec{v}}{c_\m^2} \,.
\eea
Taking into account condition~(\ref{eacc_cond_reine}) and $\hbe=\frac{m_0\,c_\m^2}{\Omega_\m}$, these equations become Eqs.~(\ref{eacc_Econc}) and (\ref{eacc_Pconc}).


\end{document}